\documentclass[runningheads]{llncs}
\usepackage{graphicx}
\usepackage{xcolor}
\usepackage{subcaption}
\begin{document}
%

\title{Decoding the visual attention of pathologists to reveal their level of expertise}

%
%
\author{Souradeep Chakraborty\inst{1} \and
Dana Perez\inst{4} \and
Paul Friedman\inst{4} \and
Natallia Sheuka\inst{4} \and
Constantin Friedman\inst{4} \and
Oksana Yaskiv\inst{4} \and
Rajarsi Gupta\inst{2} \and
Gregory J. Zelinsky\inst{3} \and
Joel H. Saltz\inst{2}\and
Dimitris Samaras\inst{1}
}
\authorrunning{Chakraborty et al.}
%
\institute{Department of Computer Science, Stony Brook University, USA
\email{\{souchakrabor,samaras\}@cs.stonybrook.edu} \and
Department of Biomedical Informatics, Stony Brook University, USA\\
\email{\{Rajarsi.Gupta,Joel.Saltz\}@stonybrookmedicine.edu} \and
Department of Psychology, Stony Brook University, USA\\
\email{\{gregory.zelinsky\}@stonybrook.edu}
\and
Department of Pathology and Laboratory Medicine, Northwell Health Laboratories, USA\\
\email{\{dperez23,pfriedman1,nsheuka,cfriedman7,OYaskiv\}@northwell.edu}
}

\maketitle              
\begin{abstract}
We present a method for classifying the expertise of a pathologist based on how they allocated their attention during a cancer reading. We engage this decoding task by developing a novel method for predicting the attention of pathologists as they read whole-slide Images (WSIs) of prostate and make cancer grade classifications. Our ground truth measure of a pathologists’ attention is the x, y and z (magnification) movement of their viewport as they navigated through WSIs during readings, and to date we have the attention behavior of 43 pathologists reading 123 WSIs. These data revealed that specialists have higher agreement in both their attention and cancer grades compared to general pathologists and residents, suggesting that sufficient information may exist in their attention behavior to classify their expertise level. To attempt this, we trained a transformer-based model to predict the visual attention heatmaps of resident, general, and specialist (GU) pathologists during Gleason grading. Based solely on a pathologist’s attention during a reading, our model was able to predict their level of expertise with 75.3\%, 56.1\%, and 77.2\% accuracy, respectively, better than chance and baseline models. Our model therefore enables a pathologist’s expertise level to be easily and objectively evaluated, important for pathology training and competency assessment. Tools developed from our model could also be used to help pathology trainees learn how to read WSIs like an expert.

\end{abstract}
\section{Introduction}

Radiology has long appreciated the role played by attention during cancer readings \cite{gandomkar2016icap,tourassi2013investigating,venjakob2012radiologists,wang2022follow}, and a similar appreciation is now growing in digital pathology   \cite{brunye2020eye,brunye2017accuracy,chakraborty2022predicting,chakraborty2022visual,sudin2021eye}. Being able to predict the visual attention of pathologists as they read whole-slide images (WSIs) will be crucial for next generation computer-assisted clinical decision support systems, but here our focus is on pathology training with respect to attention and determining whether a trainee is allocating their attention like a specialist. The studies most closely related to our goal is recent work attempting to predict the cursor-based movements of a pathologist's viewport as a measure of attention during a pathology reading  \cite{chakraborty2022predicting,chakraborty2022visual}. In \cite{chakraborty2022visual}, they did this for prostate using a fine-tuned ResNet34, and in \cite{chakraborty2022predicting} a model based on a swin-tranformer was used to predict attention during multi-stage GI-NETs examination, although the former study is most relevant to ours because they also predicted differences in attention heatmaps between genitourinary (GU) specialists and general pathologists during prostate cancer grading. However, their study was limited to only five WSIs viewed by 13 pathologists. More broadly, the scarcity of data, in terms of both the number of pathologists and WSIs, prevent discovery of patterns in a pathologist's attention behavior and severely limits the training of models to make more accurate attention predictions. More data are needed to gain deeper insights into how pathologists of different expertise allocate their attention during readings, and to develop the predictive tools that can help pathology trainees attend like specialists. 

\begin{figure}[t]
\centering
\includegraphics[width = 12.30cm]{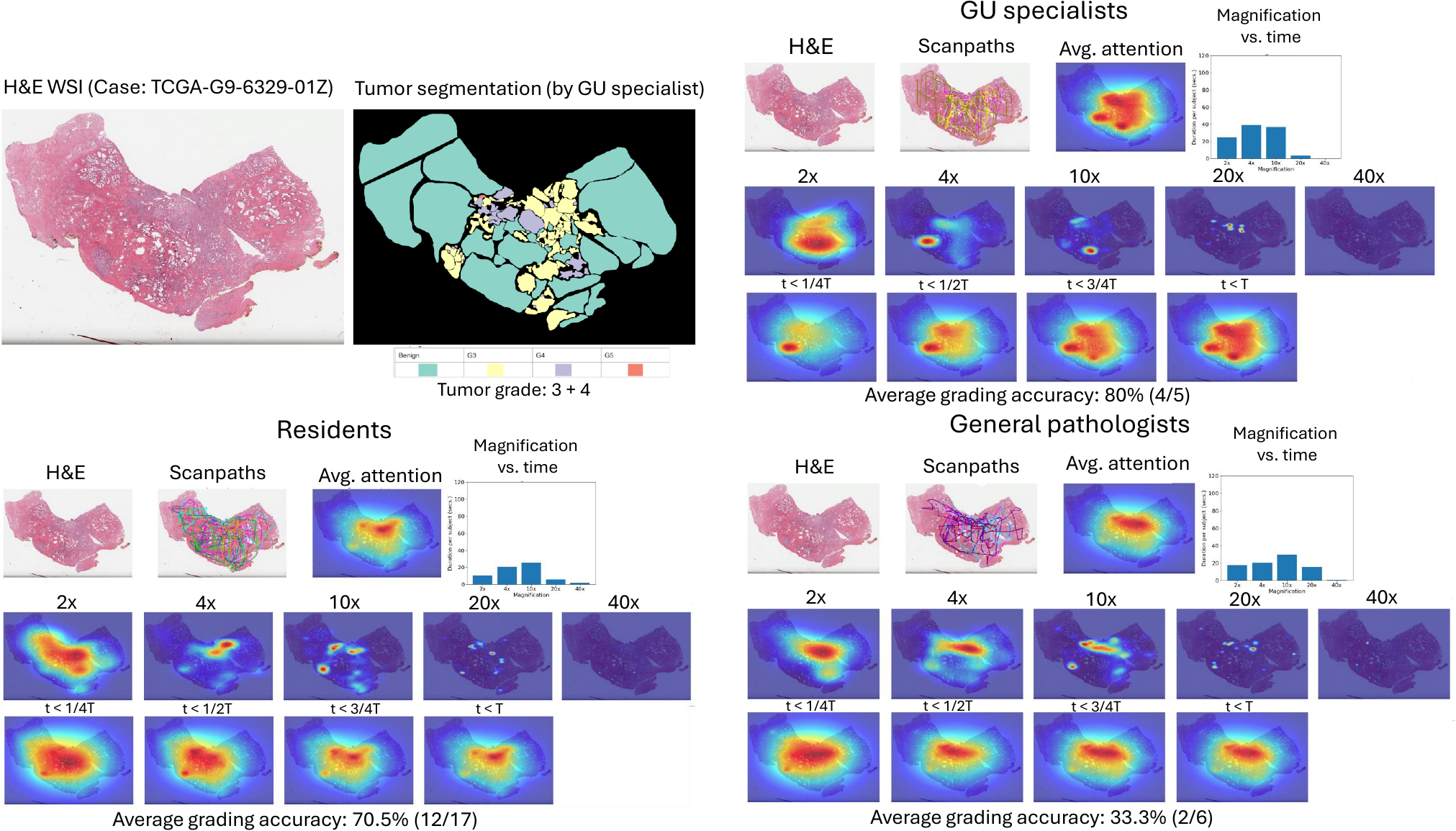}
\caption {Attention heatmaps computed for GU specialists (top-right) and general and resident pathologists (bottom). More detailed heatmaps are also shown for different levels of magnification and viewing durations. 
Left column, upper-right: grade-level segmentation of a WSI by a GU specialist. The attention heatmaps of GU specialists correlate higher with the tumor annotations compared to the non-specialists, and the specialists have the highest grading accuracy.  
}
\label{fig:teaser}
\vspace{-4mm}
\end{figure}

We help remedy this data scarcity problem by collecting the largest dataset of pathologist attention to date: 43 pathologists reading 123 WSIs of prostate cancer, yielding a total of 1016 attention trajectories. With this larger dataset, we can begin to study the relationship between a pathologist's attention and their cancer grading and how the attention of a specialist differs from that of a pathology trainee. To highlight the richness of our dataset, in Fig.~\ref{fig:teaser} we visualize attention heatmaps computed for pathologists having three levels of expertise in the pathology task: GU specialists, general pathologists, and residents. The top row for each shows the WSI with overlaid attention trajectory and an average attention heatmap, and the middle and bottom rows show more specific heatmaps for five magnification levels and for four quartiles of reading time. Qualitative differences in attention between pathologists having different levels of expertise appear almost everywhere you look, exemplifying both the richness of the data and the complexity of the task. Residents and general pathologists clearly allocated their attention to different regions in the WSI and at different magnifications,  compared to the GU specialists, factors possibly affecting grading accuracy. Also shown is a grade-level tumor segmentation of the WSI by another GU specialist, where we found a higher correlation between specialist attention and this ground truth (cross correlation = 0.452) compared to general pathologists (cross correlation = 0.418) and residents (cross correlation = 0.413).

\begin{figure}[t]
\centering
\includegraphics[width = 12.30cm]{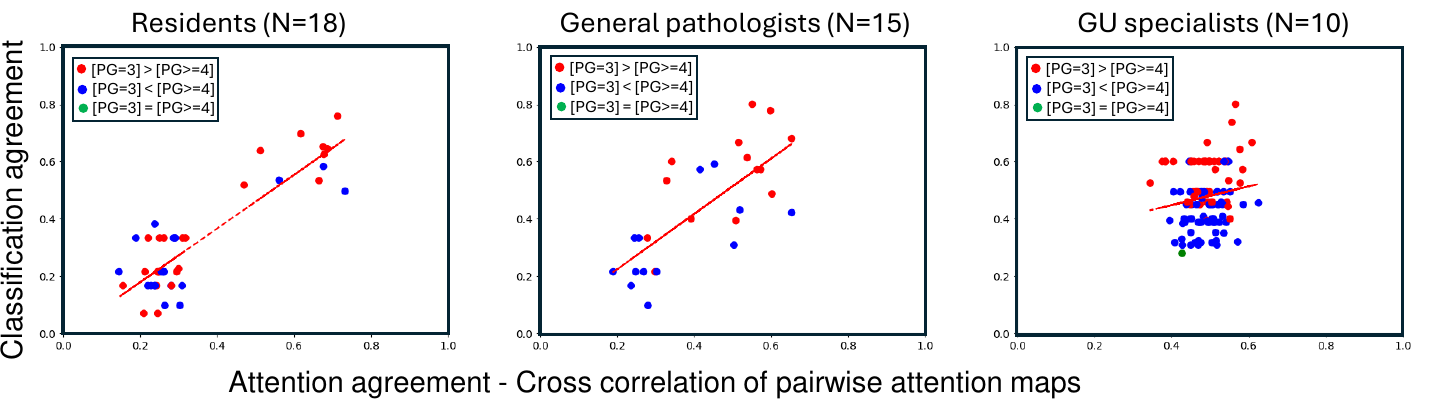}
\caption {Grade concordance vs. attention heatmap correlation across three groups of pathologists based on their expertise level. Each point represents a WSI. $PG=3$ and $PG>=4$ indicate the number of instances in which a WSI was assigned primary grade (PG) = 3 and $PG \geq 4$ respectively.
}
\label{fig:teaser2}
\end{figure}

Motivated by this analysis, we introduce ProstAttFormer, a transformer-based model designed to predict pathologists' visual attention through attention heatmaps across various magnification levels. Prior  approaches \cite{chakraborty2022predicting,chakraborty2022visual} rely on patch-wise training at a single magnification (10x) and while \cite{chakraborty2022predicting} attempted to refine the predictions by addressing spatial discontinuities caused by patch-based processing, they perform poorly due to the hand-crafted feature correspondences across patches. Our ProstAttFormer improves attention prediction performance (at different magnifications) by more effectively leveraging inter-patch feature correspondences via multi-headed self-attention mechanism in transformers, which significantly improves prediction performance. 

As a step towards developing the AI-based pathologist training tools, we develop a model that can predict pathologist expertise based on their attention behavior as they grade WSIs. The model leverages cumulative temporal attention heatmaps and magnification-wise attention heatmaps for classifying a pathologist as resident/general/specialist pathologist. 

In summary, our main contributions are: (1) the largest known pathologist attention dataset (123 WSIs across 43 pathologists), (2) a transformer-based attention prediction model that outperforms existing models, (3) a pathologist expertise prediction model based on their attention.

\section{Dataset of Pathologist Attention and Cancer Grades}
\subsection{Dataset creation}
Similar to ~\cite{chakraborty2022visual}, we used the QuIP caMicroscope, a web-based toolset for digital pathology data management and visualization \cite{saltz2017containerized} for recording the attention data of pathologists as they viewed WSIs of prostate cancer tissues (TCGA-PRAD dataset) for tumor grading. We collected attention data from 43 pathologists spanning resident (18), general (15), and GU specialist (10) levels of expertise. After reading the  instruction/consent screens, a pathologist (remotely located) was shown a WSI fit into their viewport (no magnification). They were free to navigate through the WSI in x,y,z as they conducted their reading and the GUI recorded their $1050\times1680$ viewport image at each mouse-cursor sample (20 Hz). Upon concluding their reading, the pathologist entered the tumor grade (primary and secondary) and a level of confidence in each decision on our interface. This basic procedure iterated for all the readings in the experiment. 

The 123 WSIs we used for our study were selected among 342 whole slide images of the TCGA-PRAD dataset \cite{zuley2016radiology} by a general pathologist.  In total, the data collection resulted in 1016 attention scanpaths with 329, 158 and 529 scanpaths from residents, general, and specialist pathologists respectively. On average, each WSI was examined by approximately 8 pathologists. The average viewing time per slide per pathologist was 94.68 seconds. Additionally, a GU specialist pathologist annotated the Gleason grades on a set of 22 WSIs among the 123 WSIs. We processed this  attention data in terms of attention heatmap that captures the  aggregate spatial distribution of the pathologist's attention, similar to \cite{chakraborty2022predicting,chakraborty2022visual}.  

\subsection{What is the relationship between attention during a reading and a cancer classification, and how does this change with pathologist expertise?}

Multiple factors contribute to variability in cancer diagnoses \cite{bombari2012thinking,brunye2017accuracy}, with variability in a pathologist’s attention recently added to this list \cite{chakraborty2022predicting,chakraborty2022visual}. We extend this work by characterizing in our dataset the relationship between variability in attention during cancer readings and variability in cancer classifications. 
We estimate agreement in tumor grading by computing an average pairwise concordance score as:
\begin{equation}
Conc_{Grade}^{i,j} = 1 – \frac{\sqrt{(PG_i - PG_j)^2 + (SG_i - SG_j)^2}}{\sqrt{(PG_i - PG_j)^2_{max} + (SG_i - SG_j)^2_{max}}}, 
\end{equation}
where, $Conc_{Grade}$ is the normalized score concordance between the primary and secondary Gleason scores of a pair of pathologists $i$ and $j$. Concordance scores closer to 1 indicates better agreement. We estimate agreement in attention by computing an attention heatmap for a given WSI viewing for each pathologist, then obtaining the average pair-wise cross-correlation between the different heatmaps. We hypothesize finding that variability in a non-specialists attention will lead to variability in their cancer classifications more so than specialists, who as a group will tend to agree more both on how a cancer should be graded and where they should look for it in a WSI. The pattern shown in Fig.~\ref{fig:teaser2} confirms our hypothesis. Plotted is the degree of concordance in cancer classification (y-axes) against the degree of variability in attention (x-axes) for pathologists at different expertise levels. Each data point represents a WSI examined by at least two pathologists of the same expertise level.  The average concordance scores were 0.32, 0.43, and 0.48 for residents (N=18), general pathologists (N=15), and GU specialists (N=10), respectively. Regression lines fit to these data show positive correlations; WSIs with low attention agreement tend to have low concordance scores in cancer classification, while those with high attention agreement exhibit high concordance scores. This analysis highlights a positive correlation between attention variability during WSI examination and variability in tumor grading concordance. 

Equally clear is that the strength of these correlations depends on the level of expertise. Correlations were strong and significant for residents ($r = 0.88, p < 0.01$) and general pathologists ($0.73, p < 0.01$) but weaker and not significantly different from 0 for specialists ($0.15, p = 0.09$). We interpret this expertise difference to mean that specialists tend to agree on where they should attend in a WSI and this agreed upon focus leads to greater agreement in classifications, but some resident and general pathologists (the clusters near 0.2 correlation) are still learning where to attend and consequently missing or misclassifying cancers.

\begin{figure}[t]
\centering
\includegraphics[width = 11.30cm]{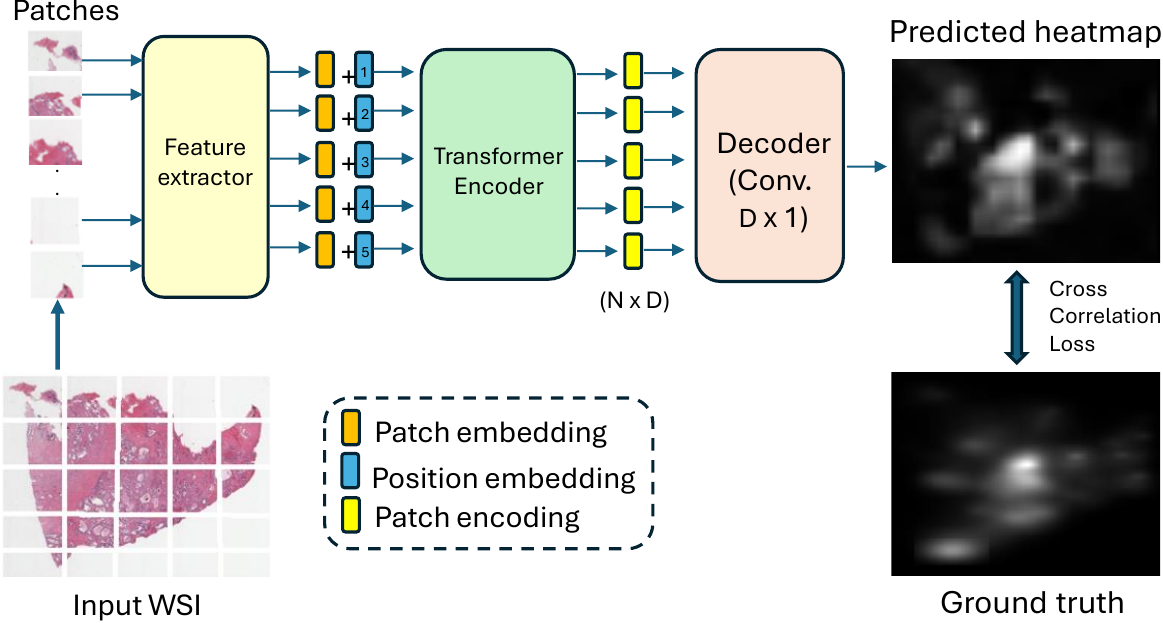}
\caption {Proposed attention prediction model ProstAttFormer that predicts pathologists attention on a WSI at different magnification levels. 
}
\label{fig:vis1}
\vspace{-4mm}
\end{figure}

\section{Methodology}

\begin{figure}[t]
\centering
\includegraphics[width = 12.30cm]{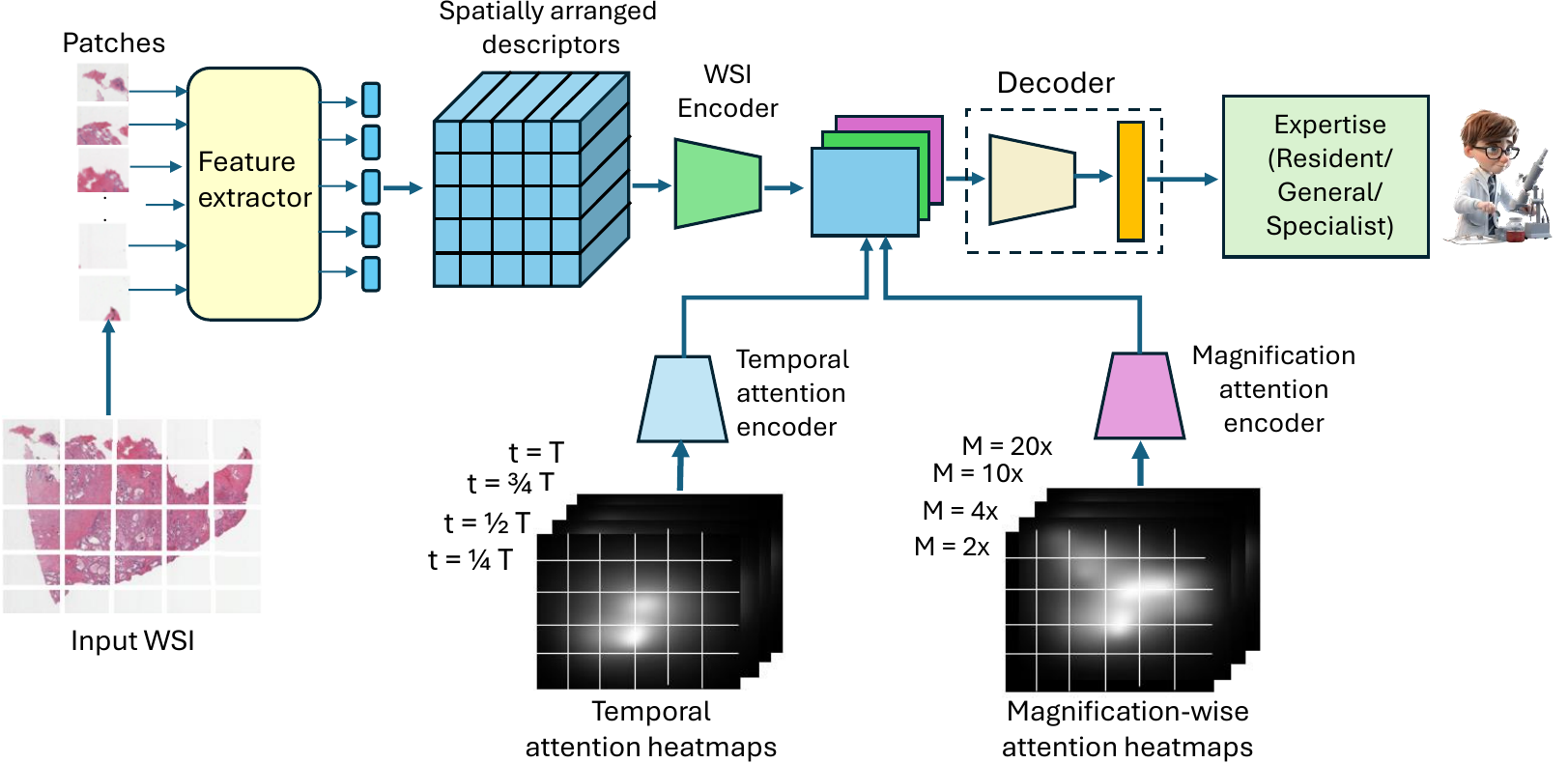}
\caption {ExpertiseNet, our pathologist expertise prediction model based on their attention. We input: (1) frozen ViT feature descriptors from a self-supervised learning model (arranged in 2D), (2) temporal attention heatmaps i.e. the cumulative attention heatmaps at different viewing durations, and (3) magnification-wise attention heatmaps to the model, which predicts the pathologist expertise. 
}
\label{fig:vis2}
\vspace{-4mm}
\end{figure}

\subsection{Predicting attention heatmaps}
Here we explain our attention heatmap prediction model ProstAttFormer in details. In Fig.~\ref{fig:vis1} we depict our ProstAttFormer model. We first split  an input WSI $I$ into a sequence of $N$  patches $I = [I_1, ..., I_N] \in R^{N \times P^2 \times C}$ where $(P, P)$ is the patch size, $N = HW/P^2$ is the number of patches (H,W are the image dimensions) and $C$
is the number of channels. Next, we extract patch-wise feature embeddings $I_0 = [F_{I_1},..,F_{I_N}] \in R^{N \times D} $ where
$F \in R^{D \times P^2}$,  using an off-the-shelf feature extractor (e.g. ResNet50 \cite{he2016deep}, DINO\cite{caron2021emerging}). To capture positional information,
learnable position embeddings
$pos = [pos_1, ..., pos_N] \in R^{N \times D}$ are
added to the sequence of patches to get the resulting input sequence of tokens
$z_0 = I_0+pos$.
A transformer \cite{vaswani2017attention} encoder composed of $L$ layers is applied to the sequence of tokens $z_0$ to generate a sequence of contextualized
encodings $z_L \in R^{N \times D}$. The sequence of patch encodings $z_L \in R^{N \times D}$ is
decoded to a heatmap $s \in R^{H \times W}$ using the decoder (a $D \times 1$ convolutional layer) that learns to map the patch-level encodings $z_L$ to patch-level attention scores. The final predicted heatmaps are obtained after map normalization. We used loss $L = 1 - CC(M_{Prd},M_{GT})$ to train this network, where, CC = cross correlation score between the predicted map $M_{Prd}$ and the ground truth map $M_{GT}$.

\subsection{Attention guided pathologist expertise prediction}
We introduce ExpertiseNet, a convolutional network that learns to predict the pathologist expertise (classification) based on how they have allocated attention across time and across different magnification levels.  Fig.~\ref{fig:vis2} outlines our model. Specifically, this model accepts: (1) frozen ViT feature descriptors from a self-supervised learning model, (2) cumulative temporal attention heatmaps i.e. the heatmaps at different viewing durations e.g. 1/4th, 1/2, 3/4th and all of the  viewing duration, (3) magnification-wise attention heatmaps (at 2x, 4x, 10x and 20x). We train this model via the weighted cross-entropy loss between the predicted logits and the ground truth labels. First, we extract feature embeddings $F_i$ for each patch $i$ in the WSI $I$ and stack them in a 3D feature embedding matrix, $F_I$. Next, we pass this  matrix through the WSI encoder while encoding the temporal and the magnification-wise attention maps using their respective encoders (each encoder block being composed of convolutional layers). The encoder outputs are stacked into a single volume and passed through a decoder to obtain the predictied labels. 

\begin{table}[htbp]
    \centering
    \caption{Comparison of the 5-fold cross validation performance on the baseline  models (colored blue) and our models (colored red) for 25 test H\&E WSIs of prostate cancer at different magnifications. We evaluate ProstAttNet \cite{chakraborty2022visual} and PathAttFormer \cite{chakraborty2022predicting} only at 10x following their original implementation.}
    \begin{minipage}[t]{0.45\textwidth}
        \centering
        \begin{subtable}{\linewidth}
            \centering
            \resizebox{6.2cm}{!}{
            \begin{tabular}{|l|l|l|l|l|l|}
        \hline
         Model & CC$_{Attn}$ & NSS$_{Attn}$ & KLD$_{Attn}$\\
        \hline
        
        \color{blue} Frozen ResNet50+Dec. \color{black}   & $0.498\pm0.214$ & $0.748\pm0.307$  & $0.383\pm0.023$\\
        \color{blue} Frozen DINO+Dec. \color{black}  & $0.486\pm0.192$  & $0.705\pm0.275$ &  $0.397\pm0.026$\\
        \color{red} Our (w/ DINO)  \color{black}  & $\textbf{0.560}\pm0.199$  & $\textbf{0.836}\pm0.290$ & $0.362\pm0.070$\\
        \color {red} Ours (w/DINO-v2) \color{black}  & $0.551\pm0.149$  & $0.829\pm0.202$  & $\textbf{0.348}\pm0.022$\\
        \hline
        \end{tabular}
        }
            \caption{2x}
        \end{subtable}
        \vspace{0.5cm} 
        \begin{subtable}{\linewidth}
            \centering
            \resizebox{6.2cm}{!}{
            \begin{tabular}{|l|l|l|l|l|l|}
        \hline
         Model & CC$_{Attn}$ & NSS$_{Attn}$ & KLD$_{Attn}$\\
        \hline
        
        \color{blue} Frozen ResNet50+Dec. \color{black}   & $0.636\pm0.067$ & $1.106\pm0.190$  & $0.512\pm0.151$\\
        \color{blue} Frozen DINO+Dec. \color{black}  & $0.595\pm0.067$  & $1.014\pm0.207$ &  $0.539\pm0.141$\\
        \color{red} Our (w/ DINO)  \color{black}  & $\textbf{0.668}\pm0.079$  & $1.175\pm0.268$ & $0.402\pm0.071$\\
        \color {red} Ours (w/DINO-v2) \color{black}  & $0.666\pm0.074$  & \textbf{1.181}$\pm0.264$  & $\textbf{0.397}\pm0.062$\\
        \hline
        \end{tabular}
        }
            \caption{4x}
        \end{subtable}
    \end{minipage}
    \hfill
    \begin{minipage}[t]{0.45\textwidth}
        \centering
        \begin{subtable}{\linewidth}
            \centering
            \resizebox{6.2cm}{!}{
            \begin{tabular}{|l|l|l|l|l|l|}
        \hline
         Model & CC$_{Attn}$ & NSS$_{Attn}$ & KLD$_{Attn}$\\
        \hline
        
        \color{blue} Frozen ResNet50+Dec. \color{black}   & $0.682\pm0.018$ & $1.510\pm0.242$  & $0.820\pm0.249$\\
        \color{blue} Frozen DINO+Dec. \color{black}  & $0.659\pm0.027$  & $1.436\pm0.236$ &  $0.860\pm0.253$\\
        \color{blue} ProstAttNet \cite{chakraborty2022visual} \color{black}  & $0.262\pm0.017$  & $0.883\pm0.138$ & $2.666\pm0.562$\\
        \color{blue} PathAttFormer \cite{chakraborty2022predicting} \color{black}  & $0.294\pm0.014$  & $0.924\pm0.145$ & $2.513\pm0.520$\\
        \color{red} Our (w/ DINO)  \color{black}  & $\textbf{0.739}\pm0.029$  & $\textbf{1.711}\pm0.360$ & $0.473\pm0.068$\\
        \color {red} Ours (w/DINO-v2) \color{black}  & $0.738\pm0.029$  & $1.710\pm0.362$  & $\textbf{0.473}\pm0.055$\\
        \hline
        \end{tabular}
        }
            \caption{10x}
        \end{subtable}
        \vspace{0.5cm}
        \begin{subtable}{\linewidth}
            \centering
            \resizebox{6.2cm}{!}{
            \begin{tabular}{|l|l|l|l|l|l|}
        \hline
         Model & CC$_{Attn}$ & NSS$_{Attn}$ & KLD$_{Attn}$\\
        \hline
        
        \color{blue} Frozen ResNet50+Dec. \color{black}   & $0.372\pm0.042$ & $ 1.910\pm0.277$  & $2.361\pm 0.503$\\
        \color{blue} Frozen DINO+Dec. \color{black}  & $0.365\pm0.062$  & $1.892\pm0.271$ &  $2.369\pm0.511$\\
        \color{red} Our (w/ DINO)  \color{black}  & $0.417\pm0.065$  & $\textbf{2.266}\pm0.368$ & $1.741\pm0.349$\\
        \color {red} Ours (w/DINO-v2) \color{black}  & $\textbf{0.419}\pm0.062$  & $2.264\pm0.377$  & $\textbf{1.731}\pm0.341$\\
        \hline
        \end{tabular}
        }
            \caption{20x}
        \end{subtable}
    \end{minipage}

\label{tab:tab1}
\vspace{-12mm}
\end{table}

\paragraph{Implementation details:} We used the ViT-S model (embedding size $D = 384$) trained using DINO for extracting the WSI patch features (frozen while training). For heatmap prediction, we input grids of variable sizes to our network for different magnifications - $10\times10$ for 2x, $20\times20$ for 4x, $50\times50$ for 10x and $60\times60$ for 20x. 
Our transformer encoder contains $L=12$ layers with $n_h = 8$ attention heads. 
For ExpertiseNet, we used a grid size corresponding to 20x magnification. The WSI encoder  consists of a \textit{conv(D,16,1)} layer (D = embedding size). The magnification and temporal attention map encoders consists of a \textit{conv(4,16,1)} layer each. The decoder consists of a \textit{AvgPool2d (k=3, stride=2)} layer followed by a \textit{conv(32,1,1)} layer and an $fc(256,3)$ layer for final class prediction. 
We used batch size = 8, learning rate = $10^{-4}$, and weight decay = $10^{-4}$, for both attention heatmap prediction and expertise prediction tasks.

\section{Results}
\label{sec:results}
\vspace{-1mm}
\subsection{Qualitative Evaluation}
\begin{figure}
\centering
\includegraphics[width = 12.30cm]{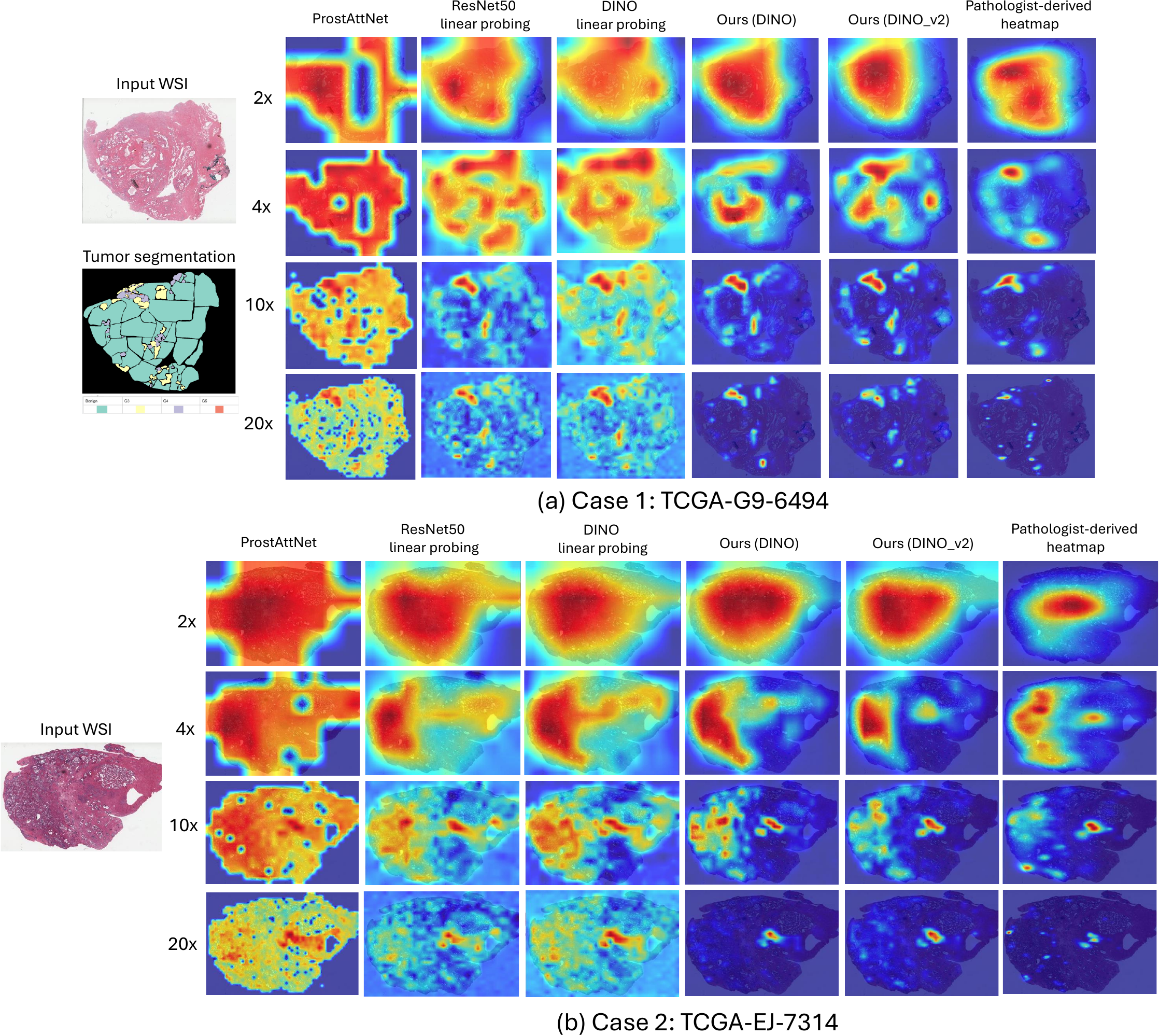}
\caption {Comparison of attention heatmap prediction performance of the proposed model ProstAttFormer with other baselines. Our ProstAttFormer model better predicts the heatmaps compared to other baselines across all magnifications for both of the two test WSI instances (cases 1 and 2).
}
\label{fig:vis3}
\vspace{-4mm}
\end{figure}

In Fig.~\ref{fig:vis3}, we qualitatively compared the attention heatmaps predicted by our model (with DINO \cite{caron2021emerging} and DINO-v2 \cite{oquab2023dinov2} features as input) with three baseline models: (1) frozen Resnet50 encoded features + linear probing using a $2048\times1$ convolutional layer as decoder, (2) frozen DINO encoded features + linear probing using a $384\times1$ convolutional layer as decoder, (3) ProstAttNet \cite{chakraborty2022visual} on two test WSI instances from our dataset. Case 1 is an WSI instance with tumor annotation from a GU specialist and case 2 is an instance without the tumor annotation. We see that our ProstAttFormer model produces more accurate attention heatmaps  compared to the baselines in both cases.  

\subsection{Quantitative Evaluation}
\vspace{-1mm}
We quantitatively evaluate the model performance using three metrics: Cross Correlation (CC), Normalized Scanpath Saliency (NSS), and KL-Divergence (KLD).  A high CC value indicates a higher correlation between the map intensities. NSS measures the average normalized attention intensity at the viewport centers and KLD measures the  dissimilarity between two probability distributions of the predicted and the ground truth attention heatmaps. 

In Tab.~\ref{tab:tab1}, we compare the 5-fold cross validation performance of the different baseline models with our models on 25 test H\&E WSIs at different magnification levels. Our models trained using the DINO and DINO-v2 feature descriptors outperform the baselines models by a significant margin by all metrics.
    
In Tab.~\ref{tab:tab2}, we compare the attention prediction performance of our ProstAttFormer model trained on specialist data with that of our model trained on non-specialist (residents and general pathologists) data. We test these models on 17 H\&E WSIs (with tumor annotations from a GU specialist) at different magnification levels. We find that our model trained on specialist data performs better than our model trained on non-specialist data on the 4x, 10x and 20x magnifications (the commonly used magnifications at which prostate cancer grading is usually done). This analysis further supports our premise that training non-specialist pathologists based on specialists' attention behavior would be effective.
\begin{table}[htbp]
    \centering
    \caption{Performance comparison of our attention prediction model trained on specialist data with a model trained on the non-specialists' (general pathologists and residents)  attention data based on attention-tumor overlap on 17 test H\&E WSIs of prostate cancer at different magnifications.}
    \vspace{0.6mm}
    \begin{minipage}[t]{0.45\textwidth}
        \centering
        \begin{subtable}{\linewidth}
            \centering
            \resizebox{6.1cm}{!}{
        \begin{tabular}{|l|l|l|l|l|l|}
        \hline
         Model & CC$_{Seg}$ & NSS$_{Seg}$ & KLD$_{Seg}$\\
        \hline
         ProstAttFormer - Specialist \color{black}  & $0.285$ & $1.032$  & $2.487$\\
        
        ProstAttFormer - Non-Specialist \color{black}  & \textbf{0.314}  & \textbf{1.027}  & \textbf{2.418}\\
        
        \hline
        \end{tabular}
        }
        \caption{2x}
        \end{subtable}
        \vspace{0.5cm} 
        \begin{subtable}{\linewidth}
            \centering
            \resizebox{6.1cm}{!}{
            \begin{tabular}{|l|l|l|l|l|l|}
            \hline
             Model & CC$_{Seg}$ & NSS$_{Seg}$ & KLD$_{Seg}$\\
            \hline
            ProstAttFormer - Specialist \color{black}  & \textbf{0.406} & \textbf{1.263}  & \textbf{2.186}\\
            
            ProstAttFormer - Non-Specialist \color{black}  & 0.386  & 1.253  & 2.250\\
            
            \hline
            \end{tabular}
        }
        \caption{4x}
        \end{subtable}
    \end{minipage}
    \hfill
    \begin{minipage}[t]{0.45\textwidth}
        \centering
        \begin{subtable}{\linewidth}
            \centering
            \resizebox{6.1cm}{!}{
            \begin{tabular}{|l|l|l|l|l|l|}
            \hline
             Model & CC$_{Seg}$ & NSS$_{Seg}$ & KLD$_{Seg}$\\
            \hline
            ProstAttFormer - Specialist \color{black}  & \textbf{0.582} & \textbf{1.851}  & \textbf{1.584}\\
            
            ProstAttFormer - Non-Specialist \color{black}  & 0.561  & 1.814  & 1.690\\
            
            \hline
            \end{tabular}
        }
        \caption{10x}
        \end{subtable}
        \begin{subtable}{\linewidth}
            \centering
            \resizebox{6.1cm}{!}{
            \begin{tabular}{|l|l|l|l|l|l|}
            \hline
             Model & CC$_{Seg}$ & NSS$_{Seg}$ & KLD$_{Seg}$\\
            \hline
            ProstAttFormer - Specialist \color{black}  & \textbf{0.592} & \textbf{2.619}  & \textbf{1.382}\\
            
            ProstAttFormer - Non-Specialist \color{black}  & 0.566  & 2.310  & 1.563\\
            
            \hline
            \end{tabular}
        }
            \caption{20x}
        \end{subtable}
    \end{minipage}
    \label{tab:tab2}
    \vspace{-10mm}
\end{table}

In Tab.~\ref{tab:expertise_prediction} we ablate the 5-fold cross validation classification performance (specialist vs. non-specialist (a) and resident/general/specialist pathologist (b) ) of our pathologist expertise prediction model ExpertiseNet using temporal attention heatmaps, magnification-wise attention heatmaps and a combination of the two heatmaps. Here, we report the classification accuracy, F1-score and the AUC score of classification with all of these higher the better. The F1 score emphasizes the balance between precision and recall, making it suitable for situations where false positives and false negatives have different costs. On the other hand, the AUC score assesses the overall discriminative ability of the model across all possible threshold settings. ExpertiseNet has the best performance when a combination of both the temporal and the magnification heatmaps is input to the model instead of separately inputting these two heatmaps.

\begin{table}[t]
    \centering
    \caption{Pathologist expertise classification performance, using temporal heatmaps, magnification-wise heatmaps and a combination of both via 5-fold cross validation on our attention dataset. Combining both information produces the best results.}
\begin{minipage}[t]{\textwidth}
        \centering
        \begin{subtable}{\linewidth}
            \centering
            \resizebox{11.5cm}{!}{
        \begin{tabular}{|l|l|l|l|l|l|}
        \hline
         Model & Accuracy & F1-score & AUC score\\
        \hline
        Random & $0.500\pm0.000$ & $0.500\pm0.000$ & $0.250\pm0.000$ \\
        ExpertiseNet (w/ Temporal heatmaps) & $0.763\pm0.013$ & $0.763\pm0.016$ & $0.764\pm0.015$ \\
        ExpertiseNet (w/ Magnification heatmaps) & $0.787\pm0.008$ & $0.785\pm0.011$ & $0.785\pm0.001$ \\
        ExpertiseNet (w/ Temporal + Magnification heatmaps) & \textbf{0.806$\pm$0.023} & \textbf{0.806$\pm$0.024} & \textbf{0.806$\pm$0.028} \\
        \hline
        \end{tabular}
        }
        \caption{2-way (specialists vs. non-specialists) cross validation results}
        \end{subtable}
    \end{minipage}
    
    \begin{minipage}[t]{\textwidth}
        \centering
        \begin{subtable}{\linewidth}
            \centering
            \resizebox{11.5cm}{!}{
            \begin{tabular}{|l|l|l|l|l|l|}
        \hline
         Model & Accuracy & F1-score & AUC score\\
        \hline
        Random & $0.333\pm0.000$ & $0.333\pm0.000$ & $0.5000\pm0.000$ \\
        ExpertiseNet (w/ Temporal heatmaps) & 0.676$\pm0.035$ & $0.630\pm0.032$ & $0.796\pm0.007$ \\
        ExpertiseNet (w/ Magnification heatmaps) & $0.731\pm0.021$ & $0.680\pm0.015$ & $0.837\pm0.020$ \\
        ExpertiseNet (w/ Temporal + Magnification heatmaps) & \textbf{0.732$\pm$0.017} & \textbf{0.696$\pm$0.010} & \textbf{0.845$\pm$0.005} \\
        \hline
        \end{tabular}
        }
        \caption{3-way (residents/general/specialist pathologists) cross validation results}
        \end{subtable}
    \end{minipage}
    \label{tab:expertise_prediction}
    \end{table}

\vspace{-1mm}
\section{Conclusion}
\label{sec:conclusion}
\vspace{-2mm}
We introduce a method to assess pathologists' expertise by analyzing their attention during cancer readings. Using a unique approach, we predict pathologists' attention while they assess prostate whole-slide images (WSIs) and classify cancer grades. By tracking their viewport movements during WSI reading, we gathered attention data from 43 pathologists over 123 WSIs. Our analysis shows that specialists demonstrate higher attention and grading agreement compared to general pathologists and residents, suggesting potential for expertise classification based on attention behavior. Employing a transformer-based model, we predict visual attention heatmaps of pathologists during Gleason grading, achieving great performance surpassing chance and baseline models. This method enables easy and objective evaluation of pathologists' expertise, crucial for pathology training and competency assessment, offering a pathway to enhance grading consensus among non-specialists by emulating AI specialists' attention patterns.

\bibliographystyle{splncs04}
\bibliography{mybibliography}

\end{document}